\documentclass[11pt]{article}
\usepackage{epsfig}

\def\Journal#1#2#3#4{{#1} {\bf #2}, #3 (#4)}

\def\NPB{{\em Nucl. Phys.} B}
\def\PLB{{\em Phys. Lett.}  B}
\def\PRL{\em Phys. Rev. Lett.}
\def\PRD{{\em Phys. Rev.} D}

\def\PRC{\em Phys. Rep.} 

\def\be{\begin{equation}}
\def\ee{\end{equation}}
\def\bea{\begin{eqnarray}}
\def\eea{\end{eqnarray}}

\begin{document}
\vspace*{2cm}
\begin{center}
{\bf EXPERIMENTAL INTRODUCTION TO EXTRA DIMENSIONS}
\footnote{Invited talk given at the session
{\it Very High Energy Phenomena in the Universe} 
of the 36th {\it Rencontres de Moriond} Les Arcs, France (January 20-27, 2001) }
\end{center}
\begin{center}
M. BESAN\c CON 
\end{center}
\begin{center}
CEA-Saclay, DAPNIA/SPP, Bat. 141. 91191 Gif sur Yvette, France\\
\end{center}
\vspace*{1cm}
\begin{center}
{\bf abstract.}\\
\end{center}
\vspace*{-0.3cm}
A short review of phenomenological and 
experimental aspects of extra spatial dimensions at colliders
is presented.\\

\section{Introduction}
How old is the idea of extra spatial dimensions ? 
The answer to this question appears to deeply vary 
wether you ask, among others, religion~\cite{kabal}, literature, philosophy, 
mathematics or physics. Focusing on physics, the first serious
discussions of extra spatial dimensions seem to bring us at the beginning
of last century with the work of Nordstr\"om~\cite{nord}, Kaluza~\cite{kalu},
Klein~\cite{klei} and then
Einstein and Bergmann~\cite{ein}
who already tackled the problem of unifying the electromagnetic
interaction with the gravitational interaction. 
\par
Although supergravity
theories formulated up to 11 spacetime dimensions and superstring theories
in 10 spacetime dimensions (10d) were known since the 70' and 80'~\cite{string},
still pursuing the goal of unifying all the known interactions, the idea of 
extra spatial dimensions received recently a new impulse. Actually,
efforts of undertanding spontaneous supersymmetry breaking 
by compactification in the context of string theories
lead already to the possibility of having extra spatial dimensions at the
TeV scale~\cite{anto1}.
Afterwards,
a better understanding of the role of branes in superstrings theories and
the relation between the five 10d superstrings theories 
in terms of duality symetries, leading to the
existence of a M-theory, having the 11d supergravity theory as its low energy
limit, has been exploited up to the striking statement that the fundamental string 
scale is viewed as an arbitrary scale which can be, formally, as low as the
TeV scale~\cite{brane} 
thus leading to the possibility of having extra spatial dimensions
at this scale. 
Futhermore, the proposal of having the standard model (SM) fields of particle
physics confined
on a 4d subspace (brane) living in a $(4+n)$d space with $n$ compact
extra spatial dimensions
where the gravitational interaction lives~\cite{add} and arguing a TeV scale as the
fundamental scale for the gravitational interaction in this $(4+n)$d space lead to
the possibility of having large compact extra spatial dimensions i.e. of the mm size 
as well as an automatic mean of solving the hierarchy problem of the SM.
It is quite remarkable that this last proposal, often referred as the ADD approach,
can be embedded within
the context of string theories~\cite{aadd}.
Even more recently, the set-up with
two 4d branes (one of which containing the SM fields)
living in a 5d space with an anti-De-Sitter (AdS) geometry provides
an additional scenario and more impulse to the idea of extra spatial dimensions. 
This approach is often referred as the model with
warped extra spatial dimension or the
Randall-Sundrum (RS) model~\cite{rs}.
\par
We don't experience more than 3 spatial dimensions in our everyday life. This 
means that extra spatial dimensions wether compact (i.e. up to 6 compact extra
space dimensions within superstring theories or 7 in M-theory) or warped (i.e.
1 within the Randall-Sundrum approach) are hidden because their sizes are
still smaller than what can be resolved by our past and past/present experimental 
apparatus. More performing experimental apparatus for new measurements of the 
gravitational law between test bodies being separated by distances below
the O(1~mm) scale where
compact extra spatial dimensions are supposed to manifest
are now being designed and/or taking data. This field of activities is described
elsewhere in these proceedings, see~\cite{fish} and~\cite{josh}.  
\par
At colliders, extra spatial dimensions, manifest themselves through
the production of Kaluza-Klein states. In the presence of an compact
extra spatial dimension $y$, a field $\phi(x_{\mu},y)$ of mass $m_o$ 
is periodic over $y$ and can be Fourier developped:
\begin{equation}
\phi(x_{\mu},y) = 
 \displaystyle \sum_{k=-\infty}^{+\infty} 
e^{{{iky}\over{R}}}
 {\phi^{(k)}(x_{\mu})}
\label{eq:kakl}
\end{equation}
where $R$ stands for the radius of the compact extra spatial dimension.
The 4d restriction ${\phi^{(k)}(x_{\mu})}$ of the field $\phi(x_{\mu},y)$
are the Kaluza Klein states (or modes or excitations) of $\phi(x_{\mu},y)$.
The number Kaluza Klein states is infinite and  Kaluza Klein states
have masses given by $m^2_k = m^2_o + ({{k^2}/ {R^2}})$.
In the following, the production and experimental signature of the various
type of Kaluza-Klein states at colliders are discussed.  
The search for these signatures and present experimental results are
discussed by E. Perez~\cite{emma}
in these proceedings. The search for extra-dimensions at future
colliders are discussed by K. Benakli~\cite{kar} also in these proceedings. 
\section{Compact extra spatial dimensions in flat geometry} 
%************************************************************************************
\subsection{TeV Gravity alone}
In the ADD approach mentioned above, only the gravitational interaction lives
in the complete space (bulk) made of the 4d subspace where the SM fields lives
and of $n$ compact extra spatial dimensions (CESD). In this approach 
the known Planck mass scale in 4d can be related to the fundamental scale 
in the bulk i.e. 
$M_{Pl(4+n)} \equiv M_D$, by:
\begin{equation}
  M^2_{Pl(4)} = M^{n+2}_{Pl(4+n)} R^n 
\label{eq:add}
\end{equation}
where $R$ is the radius of the CESD. The magnitude of the 4d Planck mass scale
is then understood as coming from a $O(1$TeV) fundamental scale $M_D$ in the 
bulk and large volumes from large CESD i.e. $R \sim 1$~mm for $n=2$
\footnote{The above relation between the
4d Planck scale and the fundamental scale in the bulk can be derived by using
the Gauss law. It can also be derived within the context of type $I/I^{\prime}$
string theory}. A $O(1$TeV) fundamental scale automatically suppresses the
hierarchy problem of the SM.
\par
The graviton is the particle associated to the gravitational interaction in the bulk.
This graviton is then a bulk graviton. The 4d SM fields couple to the 4d
restriction of the bulk graviton namely its Kaluza-Klein (KK) states. This coupling
is suppressed by the 4d Planck mass. However the smallness of this coupling is 
compensated by the mass degeneracy of KK-graviton states. The mass interval
between 2 KK-graviton states is given by 
$ \Delta m \sim ( { M_D \over TeV } )^{{n+2}\over 2} 10^{ {12n-31 \over n} } $
which gives $\Delta m \sim 3 10^{-4} eV $ for $n=2$ and $M_D=$1 TeV. 
This compensation allows for sizeable cross-sections at colliders for processes
involving the production of KK-graviton states~\cite{grav}.
\par
At colliders, the direct production of KK-graviton states will depend mainly
on the center-of-mass energies $E$ of the particles involved in the collision,
the number $n$ of CESD and the scale $M_D$ namely $ \sigma \sim {E^{n} / {M^{n+2}_D}}$.
The KK-graviton states produced are only 4d resctrictions of bulk gravitons
so that 4d KK-gravitons disappear in the bulk (from our 4d point of view).
In consequence, the direct
production of KK-graviton states at colliders gives rise to events with a 
large missing energy
component ($\not E$) in their signatures in a detector.
For example, at $e^+e^-$ colliders, KK-graviton states
can be produced in association with a $\gamma$ or a Z boson leading
$\gamma \, \not E$ or $Z \, \not E$ signatures. At $pp$ or $p \bar p$
colliders the signatures for the production of KK-graviton states
are jet $\not E$, $\gamma \, \not E$ and $Z \, \not E$.
The detection and measurement of such signatures allow for direct measurements
of the number of CESD and the scale $M_D$~\cite{grav}.
Di-fermions or di-bosons production at $e^+e^-$, $pp$, $p \bar p$ or $ep$ colliders
are also affected by processes involving KK-graviton states. These indirect effects
are signed by deviations in differential cross-sections and asymmetries measurements
with respect to the expectation from pure SM processes~\cite{grav}.
However, for $n\geq 2$,
the cross-section of these indirect processes involving KK-graviton states are 
divergent.
At the level of field theory calculations, a cut-off is usually imposed in order
to remove theses divergencies. This cut-off is unfortunately related to the
fundamental scale $M_D$ up to an arbitrary parameter usually and reasonably assumed
to be of order 1. It is worth mentioning that at the level of string theories 
calculations, in particular in the context of type I string theory, the above 
divergencies can be regularized~\cite{jiem}.
\par
Most of the searches for direct or indirect effects from large CESD
at past and past/present colliders have been performed within this ADD approach of
TeV Gravity alone. The results of these searches are discussed elsewhere in these
proceedings~\cite{emma}. The discussion of these signatures at future colliders
are also discussed elsewhere in these proceedings~\cite{kar}.
\par 
One of the most stringent constraint on $M_D$ and/or the radius $R$ of CESD comes from
the observation that KK-graviton states emission would have affected the energy
release of supernova SN1987A~\cite{hall}. 
This observation turns into the following constraints
$M_D > 50 - 130$~TeV and $R < 3 \, 10^{-4}$~mm for $n=2$. However, in the derivation
of these contraints, it is usually assumed that all large CESD radii are of the same
order of magnitude leading to an isotropy-like assumption. 
This requirement seems still to be justified~\cite{dvali}.
\par
\begin{figure}[htb]
\vspace*{5.0cm}
\begin{center}
\hspace*{-2.0cm}
\psfig{figure=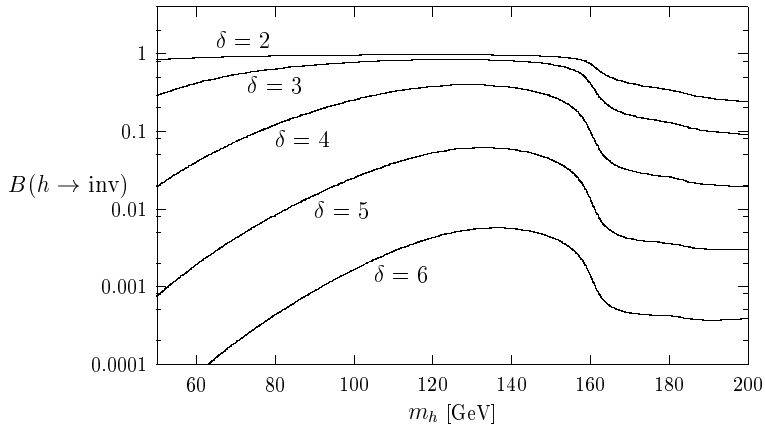,height=6.0in}
\end{center}
\vspace*{-12.0cm}
\caption{The Higgs boson invisible branching fractions as a function of its mass
for $M_D= 2$ TeV and a conformal coupling (see text) equal to 1.
\label{fig:grw}}
\end{figure}

Gravity in higher dimensional spaces does not only imply the existence
of KK-graviton states in 4 dimensions but also spin 1 and spin 0 new KK-states
which can interact with SM fields. The spin 0 states i.e. the graviscalars,
couple to the SM fields via the trace of the energy-momentum tensor. The
direct production of KK-graviscalars is suppressed relative to KK-graviton~\cite{grw}
due either
to the anomaly loop factor (trace anomaly) or to additional power of
$m^2_Z/E^2$. However, in the present field theory approach,
it is possible to consider a mixing between KK-graviscalars and the Higgs boson
which can lead to a sizeable invisible Higgs branching fraction. This invisible
Higgs branching fraction can even reach values near 1, as seen in Fig.~\ref{fig:grw}
from~\cite{grw},
depending of the conformal coupling responsible of the graviscalars-Higgs boson
mixing thus having a great impact on Higgs boson search at the Tevatron and LHC.
\par 
Before ending this section it is worth emphasizing that the ADD approach 
can be embedded within string theories and in particular the
type I string theory. More generally, when dealing with a quantum theory
of gravitation, one has always to keep in mind that string theories are 
presently the best approaches for such a theory. The spectrum of string
oscillations states does not only allow for zero mass states identified 
with the known particles but also for massive states whose masses are of the order of
the string scale. If the string scale is lowered down to a scale of O(1) TeV then
these massive states arise with the same mass and thus may contribute to observable
effects at colliders. These stringy effects can even dominate those coming from pure
KK-graviton states since their contribution to 4-points amplitudes appear as form
factors containing correction of order $g_s (E/M_s)^4$ where $g_s \sim 1/25$ and $M_s$
are the string coupling and the string scale respectively while KK-graviton effects
have $g^2_s (E/M_s)^4$ factors i.e. a $g^2_s$ dependence~\cite{jiem}.  
%************************************************************************************
\subsection{Kaluza Klein gauge bosons}
The embedding of the ADD approach within type I string theory and its brane
point of view allows to enrich the spectrum of KK states as, in this framework,
one is not only left with KK-graviton states coming from graviton in the
usual perpendicular CESD but also to the possiblity of having KK-gauge bosons 
coming from the gauge bosons of the SM living in so-called parallel CESD~\cite{acco}.
The possibility of producing KK-gauge bosons at colliders has been
actually already discussed in~\cite{vieil}.
It has also been realized that in the context of compactified type IIB string 
theory KK-excitations states having gauge interactions can arise while the
gravity becomes strong at scales kept at $O(10^{9})$ TeV~\cite{boris}. 
\par
The effects of 
KK-gauge bosons can be either seen by their effects on electroweak observables
in precision measurements or in particles production at colliders.
The analysis of the effects of KK-gauge bosons on electroweak (EW) observables often 
requires some generic assumptions such as, 1) the non influence of gravitational 
effects, 2) only one extra dimension usually compactified on the $S^1/Z_2$ orbifold,
where the $Z_2$ symmetry appears to be useful to introduce fermions chirality on
the 4d branes
localized at the fixed points of the orbifold,
3) the choice of the reference model i.e. SM, MSSM or even NMSSM and finally 4)
the localization of the fields i.e. for the SM, the fermions on the 4d branes localized
on the fixed point of the above orbifold, the gauge fields in the 5d bulk
and the Higgs field either in the brane or in the bulk. In addition, the
effective 5d gauge coupling ${\hat {g}}$ is often given in terms of the 4d gauge
coupling
$g$ i.e. ${\hat {g}}^2 \sim g^2 R$ where $R \sim 1/ M_c$ is the radius of the 
parallel CESD and $M_c$ the scale of this parallel CESD. This effective
gauge coupling in 5d has been shown to be finite while for more than one parallel CESD
the effective gauge coupling is divergent. However, in the context
of string theories, the brane configuration has to be taken into
account in order to define the gauge coupling which can then be regularized. 
The results in terms of constraints on $R$ or $M_c$ from EW precision measaurement
is given elsewhere in these proceedings~\cite{emma}. In order to fix the order
of magnitude on $M_c$, a global fit from the EW precision measurements from the LEP
experiments allows to derive $M_c > 3.5$ TeV~\cite{riwe}. 
\par
At this stage, it is important to note that 
grand unification at intermediate mass scales through extra dimensions has 
been discussed in~\cite{ddg} as early as the ADD scenario.
This analysis involves the MSSM as the reference model, and it has been shown
that the gauge couplings unification might be brought down to low energy scales
due to the presence of KK-states, including  KK-gauge bosons.
These KK-states are responsible for a power law contribution to the running
of the gauge couplings.
\par
At colliders, KK-gauge bosons can be directly produced as resonances
if their masses are kinematically accessible. The KK-gauge bosons decay
into pairs of leptons or into pairs of quarks giving rise to 2 hadronic jets.
The masses of these KK-gauge bosons are then given by the 2-leptons invariant mass
(or transverse mass) or by the 2-jets invariant mass. If their masses are 
not kinematically accessible, the effects of KK-gauge bosons is signed by
deviations in differential cross-sections and asymmetries measurements
with respect to the expectation from pure SM processes. Moreover, the clean
environment of leptonic colliders allows for a measurement of the
KK-gauge bosons coupling to fermions thus allowing for a possible model 
disentangling~\cite{rizzo}.
The perspectives for direct or indirect signals for KK-gauge bosons at
future colliders are given elsewhere in these proceedings~\cite{kar}. 
%************************************************************************************
\section{Warped extra spatial dimension}
Another approach for extra spatial dimensions has been proposed in~\cite{rs}.
In this scenario two 4d branes with tensions $V$ and $V^{\prime}$ 
are situated at $y=0$ and $y=\pi r_c$ of a 5d bulk with cosmological constant
$\lambda$ where gravitation lives. With this setup,
the metric $ds^2 = e^{-2k|y|} \eta_{\mu\nu}dx^{\mu}dx^{\nu} + dy^2$, 
where k is a scale factor of the order of the 4d Planck scale, is a solution
of Einstein equations provided $V=V^{\prime}=24\, M^3_5k$ and 
$\Lambda = - 24 \, M^3_5k^2$ i.e. a negative cosmological constant in the bulk
thus corresponding to an Anti-De Sitter (AdS) geometry. The factor  
$e^{-2k|y|}$ which is in front on the usual 4d part of the metric and which
depends on the 5th dimension is often referred as the warp factor.
\par
One of the interesting consequence of this approach comes from the observation that
a fundamental mass scale on the brane at $y=0$ is red-shifted by this warp factor
on the other brane at $y=\pi r_c$. Thus, with $k r_c \sim 12 $ a O(1) TeV mass scale
can be produced from the Planck mass scale which can provide a hint for the
understanding of the hierarchy between the EW scale and the 4d Planck mass scale.
%Reversely, the 4d Planck mass scale can be generated from a fundamental
%$O(1)$~TeV mass scale $M$.
\par
In constrast to the ADD approach, the 4d Planck mass is now given by:
\begin{equation}
 {\bar M}^2_{Pl} = { M^3_5 \over k } [ 1 - e^{-2kr_c\pi} ]  
\label{eq:rspl}
\end{equation}
which remains well defined
even for $r_c \rightarrow \infty$.
Furthermore, also in constrast to the ADD approach, 
the KK expansion of the graviton is now given in terms of linear combinations
of Bessel functions and thus the masses of KK-graviton, expected to be $O(1)$~TeV,
are no longer equally spaced but are then given by $ m_n = x_n k e^{-k \pi r_c} $ 
where $x_n$ are roots of Bessel functions.  
The coupling of the KK-graviton 0-mode state to the fields of the SM
is suppressed by the 4d Planck mass. Nevertheless, the non-zero modes
can be directly produced at colliders if kinematically accessible since their
coupling to the SM fields is only suppressed by the 4d Planck mass red-shifted by
the warp factor i.e.  $1 / (e^{-k\pi r_c} M_{Pl})$. The phenomenology of 
warped extra dimension (WED) 
usually depends on 2 parameters  $e^{-k\pi r_c}$ and $k/M_{Pl}$.
\par
\begin{figure}[htb]
\begin{center}
\hspace*{2.0cm}
{\rotatebox{90}{\psfig{figure=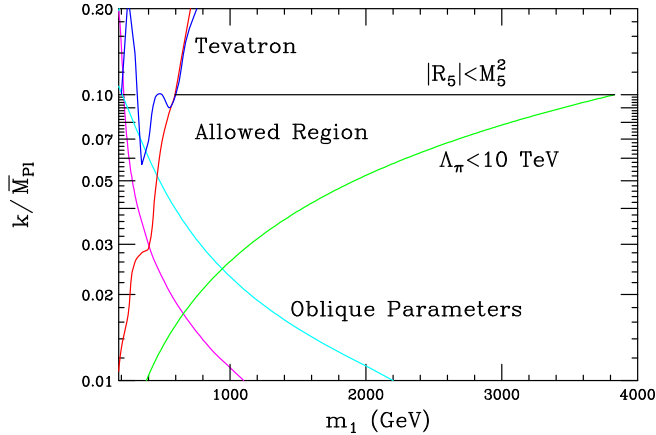,height=3.0in}}}
\caption{Allowed region in the $k/{\bar M}_{Pl}$ and
$m_1$ plane where $m_1$ is the mass of the 1st KK-graviton and 
${\Lambda_{\pi} = {\bar M}_{Pl}  e^{-kr_c\pi}}$. The oblique parameters lines come
from a global fit to the S and T oblique parameters
.}
\label{fig:rs}
\end{center}
\end{figure}

At colliders such as the Tevatron or the LHC, KK-graviton states from WED
can be produced as resonances. These resonances then decay
predominantly into two hadronic jets~\cite{hew} and this channel
channel dominates the other channels i.e. $W^+W^-$, $ZZ$, $l^+l^-$, $t\bar t$
and $hh$. Although not the dominant channel, 
the two leptons channel $l^+l^-$ allows for a clean signature of the KK-graviton
from WED at hadronic colliders such as the LHC. Then the measurement of the two leptons 
invariant mass allows to measure the mass of this KK-graviton and the measurement 
of the (polar) angular differential cross-section allows to establish its spin.
Fig.~\ref{fig:rs} from~\cite{hew} shows the allowed region for the 1st KK-graviton
state of mass $m_1$ and with 
$\Lambda_{\pi} = {\bar M}_{Pl}  e^{-kr_c\pi}$. The oblique parameters lines
come from a global fit to the S and T oblique parameters~\cite{pes}.
%************************************************************************************
\subsection{Radion phenomenology} 
\begin{figure}[htb]
\begin{center}
\epsfig{figure=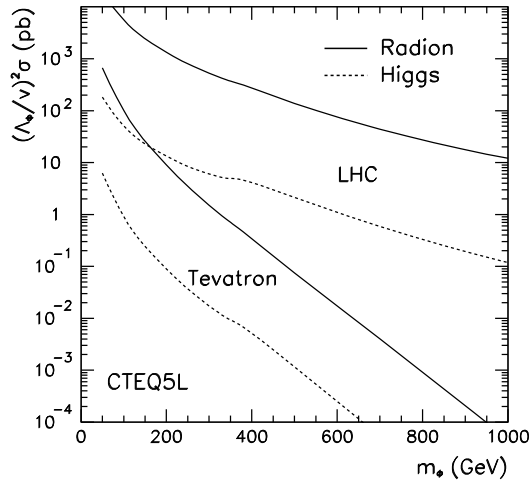,height=3.0in}
\end{center}
\caption{The radion production cross section via
gluon fusion at the Tevatron ($\sqrt {s} = 2$~TeV) and the LHC ($\sqrt {s} = 14$~TeV)
with a scale factor ($\Lambda_{\phi} / v$  where $v$ is the Higgs boson v.e.v
and $\Lambda_{\phi}$ is defined in the text)
compared to the Higgs boson  production cross sections (dashed lines).
\label{fig:ko}}
\end{figure}

In this WED approach, $r_c$ is associated with the vacuum expectation value
of a massless 4d scalar field which is known as the modulus field or the radion.
The presence of a scalar field in the bulk with interaction terms localized
on the branes, allows to stabilize the value of $r_c$~\cite{gw}. In order to have
$k r_c \sim 12 $ as argued in the previous section, the radion, after stabilization,
should be lighter than the KK-graviton states from WED and is then likely to be the
first state accessible at colliders. The radion couples to the SM fields via
the trace of the energy-momentum tensor with strength given by $1/\Lambda_{\phi}$
with $\Lambda_{\phi}=(\sqrt {24 M^3_5/k})  e^{-kr_c\pi}$. 
Fig.~\ref{fig:ko} from~\cite{ko} shows the radion production cross section via
gluon fusion at the Tevatron ($\sqrt {s} = 2$ TeV) and the LHC ($\sqrt {s} = 14$ TeV)
compared to the Higgs production cross sections.
The radion decays predominantly into 
$W^+W^-$, $ZZ$, $hh$, $t \bar t$ if kinematically allowed,
otherwise it decays mainly into a pair of gluons
and, to a less extent, into $b \bar b$. 
The radion phenomenology is very similar to the SM Higgs boson except that its
coupling to two gluons (production and decay) is enhanced by the trace anomaly.
However, it is possible to consider a possible mixing between the Higgs boson and the
radion~\cite{grw} giving rise to two new eigenstates. These new eigenstates
can have quite different branching
fractions i.e. up to factors of order 50, in particular for the decays 
into $W^+W^-$ and $ZZ$ depending on the conformal coupling
which is responsible of this Higgs-radion mixing.
\par
%************************************************************************************
\section{Conclusions} 
The recent new impulse given to the idea of extra spatial dimensions
have led to a rich spectrum of approaches either in the flat geometry stream (ADD)
and its embedding within stringy scenarios
or in the warped geometry (RS) stream. The phenomenology connected to these various
approaches is in its infancy and is still developping. These phenomenological
developpements are carried out in connection with more fundamental/theoretical
works in the hope to extract more motivated models.
Many tests can already be
performed at present and future experiments including sub-millimeter gravity
measurement and experiments at present and future colliders.
A short review of phenomenological and 
experimental aspects of extra spatial dimensions at colliders
has been presented. However many topics not covered here are highly worth to be 
looked at. This includes topics such as fermions masses within branes worlds,
phenomenology from the supersymmetrization of extra dimension models, EW and
supersymmetry breaking within brane worlds not to speak about the impact of
extra dimensions on astrophysics and cosmology.
%************************************************************************************

\section*{Acknowledgments}
It is a pleasure to thanks the organizers of the session
{\it Very High Energy Phenomena in the Universe} 
of the 36th {\it Rencontres de Moriond} 
and in particular Jacques Dumarchez for the invitation to give an experimental
introduction on extra-dimensions. It is also a pleasure to thank K.~Benakli, 
E.~Fishbach, J.~Long and E.~Perez who have participated to
this session for sharing a deep enthusiasm for the physics of
extra-dimensions.


\section*{References}
\begin{thebibliography}{99}
\bibitem{kabal} For example, exegetic considerations 
of the first hebrew word of the torah with its letters turned
into {\it beshe'erit} in place of the known {\it bereshit}  
may be difficult to date but can lead to interesting discussion
to this respect.
\bibitem{nord} G.~Nordstr\"om, Phys. Z. {\bf 15}, 504 (1914). 
\bibitem{kalu} T.~Kaluza, Sitzungsber. Preuss. Akad. Wiss. Berlin,
Math. Phys. K1, 966(1921).
\bibitem{klei} O.~Klein, {\em Z. Phys.} {\bf 37}, 895 (1926) and
Nature {\bf 118}, 516 (1926).
\bibitem{ein} A.~Einstein and P.~Bergmann, Ann. Math. {\bf 39}, 683 (1938).
\bibitem{string} The list of references concerning supergravity theories and
strings theories would be too long to be reproduced here in extenso
and we refer the reader to:\\ 
P.~van~Nieuwenhuizen, Supergravity, \Journal{\PRC}{68}{189}{1981},
M.B.~Green, J.H.~Schwartz and E.~Witten, Superstring Theory, Cambridge University
Press (1987),
J.~Polchinski, String Theory Cambridge University Press (1998),
D.~Bailin and A.~love, Supersymmetric Gauge Field Theory and String Theory,
Bristol (1994) and references therein.
\bibitem{brane}
Again, the list of references concerning
branes and dualities would be too long to be reproduced here 
and we refer the reader to:\\ 
J.H.~Schwarz,  
{\em Nucl.Phys.} Proc.Suppl. ({\bf 55B}), (1) (1997) [hep-th/960720],
J.~Polchinski, hep-th/9611050,
E.~Witten, \Journal{\NPB}{443}{85}{1995} [hep-th/9503124],
J.~Polchinski, E. Witten, 
\Journal{\NPB}{460}{525}{1996} [hep-th/9510169],  
J.~Lykken, \Journal{\PRD}{54}{3693}{1996} [hep-th/9603133],
C.~Vafa, hep-th/9702201,
A.~Sen, hep-th/9802051,
C.P.~Bachas, hep-th/9806199,
I.~Antoniadis, H.~Partouche, T.R,~Taylor, 
{\em Nucl.Phys.} Proc.Suppl. ({\bf 61A}), (58) (1998) [hep-th/9706211],
I.~Antoniadis, hep-th/0102202
and references therein
\bibitem{anto1} I.~Antoniadis, \Journal{\PLB}{246}{377}{1990}.
\bibitem{add} N.~Arkani-Hamed, S.~Dimopoulos and G.~Dvali,
\Journal{\PLB}{429}{263}{1998} [hep-ph/9803345] 
\bibitem{aadd} I.~Antoniadis, N.~Arkani-Hamed, S.~Dimopoulos and G.~Dvali,
\Journal{\PLB}{436}{257}{1998} [hep-ph/9804398], G.~Shiu and S.H.~Tye
\Journal{\PRD}{58}{106007}{1998} [hep-th/9805157], I.~Antoniadis, C.~Bachas and
E.~Dudas, \Journal{\NPB}{560}{93}{1999} [hep-th/9906039], G.~Aldazabal, L.E.~Ibanez 
and F.~Quevedo, {\em JHEP} {\bf 0001}, (031), (2000) [hep-th/9909172], 
I.~Antoniadis and K.~Benakli, {\em Int. J. Mod. Phys.} ({\bf A15}), (4237) (2000)  
[hep-ph/0007226]
and references therein
\bibitem{grav} G.F.~Giudice, R.~Rattazzi and J.D.~Wells,
\Journal{\NPB}{544}{3}{1999} [hep-ph/9811291] see also
E.A.~Mirabelli, M.~Perelstein and M.E.~Peskin, 
\Journal{\PRL}{82}{2236}{1999} [hep-ph/9811337],
T.~Han, J.D.~Lykken and R.~Zhang, \Journal{\PRD}{59}{105006}{1999}
[hep-ph/9811350], J.L.~Hewett, \Journal{\PRL}{82}{4765}{1999} [hep-ph/9811356].
\bibitem{acco} 
E.~Accomando, I.~Antoniadis and K.~Benakli,
\Journal{\NPB}{579}{3}{2000} [hep-ph/9912287].
\bibitem{vieil} 
I.~Antoniadis and K.~Benakli, 
\Journal{\PLB}{326}{69}{1994} [hep-th/9310151],
I.~Antoniadis, K.~Benakli and M.~Quiros
\Journal{\PLB}{331}{313}{1994} [hep-ph/9403290]
\bibitem{boris} I.~Antoniadis and B.~Pioline,
\Journal{\NPB}{550}{41}{1999} [hep-th/9902055].
\bibitem{riwe} T.G.~Rizzo and J.~Wells, \Journal{\PRD}{61}{016007}{2000}
[hep-ph/9906234].
\bibitem{ddg} K.R~Dienes, E.~Dudas and T.~Gherghetta, 
\Journal{\NPB}{537}{47}{1999}
[hep-ph/9806292].
\bibitem{rizzo} T.G.~Rizzo, \Journal{\PRD}{61}{055005}{2000} [hep-ph/9909232].
\bibitem{rs} L.~Randall and R.~Sundrum \Journal{\PRL}{83}{3370}{1999} and
\Journal{\PRL}{83}{4690}{1999}.
\bibitem{fish} E.~Fischbach, these proceedings.
\bibitem{josh} J.~Long, these proceedings.
\bibitem{emma} E.~Perez, these proceedings.
\bibitem{kar} K.~Benakli, these proceedings.
\bibitem{jiem} E.~Dudas and J.~Mourad, 
\Journal{\NPB}{575}{3}{2000} [hep-th/9911019],
E.~Accomando, I.~Antoniadis and K.~Benakli,
\Journal{\NPB}{579}{3}{2000} [hep-ph/9912287],
S.~Cullen, M.~Perelstein and M.E.~Peskin, 
\Journal{\PRD}{62}{055012}{2000} [hep-ph/0001166].
\bibitem{hall} L.J.~Hall and D.~Smith, \Journal{\PRD}{60}{085008}{1999} see
also I.~Antoniadis and K.~Benakli hep-ph/0004240. 
\bibitem{dvali} private communication.
\bibitem{hew} H.~Davoudiasl, J.L.~Hewett and T.G.~Rizzo,
\Journal{\PRD}{63}{075004}{2001} [hep-ph/0006041]
\bibitem{pes} M.E.~Peskin and T.~Takeuchi, \Journal{\PRL}{65}{964}{1990}
and \Journal{\PRD}{46}{1992}{381}.
\bibitem{gw} W.D.~Goldberger and M.B.~Wise,
\Journal{\PRD}{60}{107505}{1999} [hep-ph/9907218],
\Journal{\PRL}{83}{4922}{1999} [hep-ph/9907447] and
\Journal{\PLB}{475}{275}{2000} [hep-ph/9911457].
\bibitem{ko} S.B.~Bae, P.~Ko, H.S.~Lee and J.~Lee,
\Journal{\PLB}{487}{299}{2000} [hep-ph/0002224], see also
U.~Mahanta, \Journal{\PLB}{480}{176}{2000} [hep-ph/0002049] and
\Journal{\PLB}{483}{196}{2000} [hep-ph/0002183].
\bibitem{grw} G.F.~Giudice, R.~Rattazzi and J.D.~Wells, 
\Journal{\NPB}{595}{250}{2001} [hep-ph/0002178].  
\end{thebibliography}
\end{document}